\documentclass[sigconf,nonacm]{acmart}
\AtBeginDocument{%
  }

\setcopyright{none}
\copyrightyear{2027}

\usepackage{booktabs} 
\usepackage{multirow} 
\usepackage{graphicx} 
\usepackage{amsmath}  
\usepackage[dvipsnames, table]{xcolor} 
\usepackage{tabularx}
\usepackage{threeparttable}
\usepackage{microtype}
\usepackage{fontawesome5}
\renewcommand\footnotetextcopyrightpermission[1]{}

\begin{document}

\title{DocFS: File Systems as Agentic Reasoning Environments for Complex Multimodal Document Question Answering}

\author{Bingdian Zhang}
\affiliation{%
  \institution{State Key Laboratory of AI Safety}
  \city{Beijing}
  \country{China}}
\affiliation{%
  \institution{Institute of Computing Technology, Chinese Academy of Sciences}
  \city{Beijing}
  \country{China}}
\affiliation{%
  \institution{School of Computer Science and Technology, University of Chinese Academy of Sciences}
  \city{Beijing}
  \country{China}}
\email{zhangbingdian26z@ict.ac.cn}

\author{Zixuan Li}
\correspondingauthor
\affiliation{%
  \institution{State Key Laboratory of AI Safety}
  \city{Beijing}
  \country{China}}
\affiliation{%
  \institution{Institute of Computing Technology, Chinese Academy of Sciences}
  \city{Beijing}
  \country{China}}
\affiliation{%
  \institution{School of Computer Science and Technology, University of Chinese Academy of Sciences}
  \city{Beijing}
  \country{China}}
\email{lizixuan@ict.ac.cn}

\author{Long Bai}
\correspondingauthor
\affiliation{%
  \institution{State Key Laboratory of AI Safety}
  \city{Beijing}
  \country{China}}
\affiliation{%
  \institution{Institute of Computing Technology, Chinese Academy of Sciences}
  \city{Beijing}
  \country{China}}
\affiliation{%
  \institution{School of Computer Science and Technology, University of Chinese Academy of Sciences}
  \city{Beijing}
  \country{China}}
\email{bailong@ict.ac.cn}

\author{Xiaolong Jin}
\correspondingauthor
\affiliation{%
  \institution{State Key Laboratory of AI Safety}
  \city{Beijing}
  \country{China}}
\affiliation{%
  \institution{Institute of Computing Technology, Chinese Academy of Sciences}
  \city{Beijing}
  \country{China}}
\affiliation{%
  \institution{School of Computer Science and Technology, University of Chinese Academy of Sciences}
  \city{Beijing}
  \country{China}}
\email{jinxiaolong@ict.ac.cn}

\author{Jiafeng Guo}
\affiliation{%
  \institution{State Key Laboratory of AI Safety}
  \city{Beijing}
  \country{China}}
\affiliation{%
  \institution{Institute of Computing Technology, Chinese Academy of Sciences}
  \city{Beijing}
  \country{China}}
\affiliation{%
  \institution{School of Computer Science and Technology, University of Chinese Academy of Sciences}
  \city{Beijing}
  \country{China}}
\email{guojiafeng@ict.ac.cn}

\author{Xueqi Cheng}
\affiliation{%
  \institution{State Key Laboratory of AI Safety}
  \city{Beijing}
  \country{China}}
\affiliation{%
  \institution{Institute of Computing Technology, Chinese Academy of Sciences}
  \city{Beijing}
  \country{China}}
\affiliation{%
  \institution{School of Computer Science and Technology, University of Chinese Academy of Sciences}
  \city{Beijing}
  \country{China}}
\email{cxq@ict.ac.cn}

\renewcommand{\shortauthors}{Zhang et al.}

\begin{abstract}
  VLM-based agents solve complex tasks by iteratively interacting with environments, and their effectiveness depends heavily on how the environment exposes information and provides tools for acting on it. However, constructing an effective environment for the task of complex multimodal document question answering remains challenging. Documents usually span many pages with multimodal contents distributed across hierarchical sections and linked by cross-modal associations, requiring the environment to faithfully organize documents and provide agents with tools aligned with them. Thus, existing methods organize documents into structured representations, but these address only the information side of the environment, lacking tools aligned with the structure and thus limiting agents' ability to fully exploit the designed structure. We observe that file systems naturally combine structure organization with aligned tools: directories and files encode document structure, while file operations provide rich and powerful tools to access file contents. Building on this insight, we propose DocFS, a file-system-based agentic framework that reorganizes raw documents as file systems where directories mirror the section hierarchy and explicit links connect text with figures, tables, and equations. Based on the well-established file system tools such as navigation, search, and reading, agents can iteratively explore the organized document structure, locate relevant evidence, and understand fine-grained local information, thereby effectively exploiting the rich structural and semantic information in complex documents. Paired with Qwen3.5-Plus, DocFS achieves state-of-the-art performance on MMLongBench-Doc and FinRAGBench-V, outperforming the strongest baselines by 2.9 and 4.3 percentage points, respectively. Further analysis shows that DocFS is up to 10.3$\times$ faster and consumes up to 10.1$\times$ fewer tokens than the strongest baseline during reasoning.
\end{abstract}

\ccsdesc[500]{Information systems~Information retrieval}

\keywords{Complex Multimodal Documents, Document Organization, File System, Agentic Question Answering, Evidence Retrieval}


\maketitle

\section{Introduction} 

Vision language model (VLM)-based agents have demonstrated promising capabilities in solving complex tasks by iteratively interacting with external environments through environment-provided tools. Their effectiveness therefore depends not only on the reasoning capabilities of the underlying models but also on how the environment exposes information and provides tools to access it. Despite the success of this paradigm across various tasks, applying it to complex multimodal document question answering (DocQA) remains challenging. This task requires agents to locate, understand, and integrate evidence scattered across lengthy documents where multimodal contents such as text, figures, tables, and equations are distributed across hierarchical sections and linked by cross-modal associations. Consequently, effective agent-based DocQA requires an environment that organizes the document structure and provides tools aligned with it for exploration of the document content.

We view DocQA as a problem of constructing a usable reasoning environment for an agent, which requires jointly considering how document structure is represented and how the agent accesses and explores it. From this perspective, existing methods can be primarily divided into two categories according to how they organize document content: flat document organization and structured document organization. The former encodes documents primarily as page-level visual representations or flat content units, supporting evidence acquisition through retrieval or iterative exploration of pages and retrieved results \cite{faysse2025colpali, ma2024unifying, yu2025visrag, wu2025doc, jain2025simpledoc, yu2026visual, han2025mdocagent, zhu2026doclens}. Although such methods may preserve visual layout or support multi-step evidence acquisition, they do not explicitly expose document hierarchies and cross-modal relations as structures that agents can directly explore. The latter makes document structure explicit through content trees, multi-granularity indexes, page-context graphs, or other hierarchical representations \cite{wang2025bookrag, gong2026mhier, wu2025molorag, sun2025docagent, wei2026docatlas}. However, these structures serve different purposes: some primarily support retrieval, with the organized information ultimately delivered to the agent as retrieved evidence, while others expose document structures to agents through specialized operation interfaces. Consequently, explicit structure does not necessarily translate into direct and flexible access during reasoning, and interfaces designed for particular representations may require substantial system-specific customization. These limitations reveal a gap between document structure representation and agent access: existing methods either fail to make complex document structures explicitly accessible or bind access operations to specific organizational designs. A unified environment that both represents document structure and provides directly usable operations for agents to explore it therefore remains needed.

Therefore, an important design question remains: \textbf{how can a document be organized into an effective agentic reasoning environment that not only explicitly captures the document structure, but also remains directly accessible and effectively exploitable through aligned tools throughout the agent's interaction with it?} During reasoning, the agent utilizes three types of information from the document: the section hierarchy which structures the content so that agents can form a global view of the document and navigate toward candidate evidence; heterogeneous content elements, including text, figures, tables, and equations, which differ in form and require dedicated tools to access and understand; and cross-modal associations which relate elements of different modalities. To obtain this information, the agent must be able to perform the corresponding operations: navigating the section hierarchy to move between the global organization and local details, searching by keywords to locate candidate evidence, and reading files to access and understand elements of each type. We observe that a file system combines an explicit organization of information with standard operations aligned with it: nested directories mirror the section hierarchy, element-type files hold heterogeneous content, and index and relation files expose navigation paths and cross-modal links; more importantly, it provides the standard operations that agents can use directly to support the effective access to the organized structure, with no specialized interfaces to design. It is therefore a natural platform for constructing the agent's environment.

\textbf{Motivated by this insight, we propose DocFS, a file-system-based agentic framework for complex multimodal document question answering.} DocFS transforms a raw document into a hierarchical file system that reconstructs the document's section hierarchy as nested directories, stores textual content, figures, tables, and equations as element-type files in the corresponding directories, and exposes cross-modal associations explicitly. We further enhance the file system with two kinds of description files: structure description files (index and relation files) created at both the global and local levels, which turn the section hierarchy into navigation entry points and record cross-sectional and cross-modal relationships; and image description files, which make visual content retrievable by keyword through textual descriptions explicitly mapped to the original images, while the originals are retained for fine-grained visual inspection. A VLM-based agent then explores this file system through standard operations: it first navigates the hierarchy and searches for candidate evidence, then reads the relevant files and follows explicit relations to gather complementary information, and finally refers to the source files to verify details when necessary. As a result, the agent can answer a question from evidence that is structurally located, cross-modally complete, and source-verified.

We evaluate DocFS on two challenging benchmarks, MMLongBench-Doc \cite{ma2024mmlongbenchdocbenchmarkinglongcontextdocument} and FinRAGBench-V \cite{zhao-etal-2025-finragbench}, using two frontier backbone models. Paired with Qwen3.5-Plus \cite{qwen35blog}, DocFS achieves state-of-the-art performance on both benchmarks, outperforming the strongest baselines by 2.9 and 4.3 percentage points, respectively. Further analyses show that each component of the file system contributes to the final performance, and that DocFS is 4.3--10.3$\times$ faster and consumes 1.5--10.1$\times$ fewer tokens than the strongest baseline during reasoning.

In summary, our contributions are as follows:
\begin{itemize}
    \item We introduce a file-system-based representation that mirrors document section hierarchy, preserves heterogeneous content elements, and explicitly exposes cross-modal associations.
    \item We develop an agentic framework that leverages standard file system operations to navigate organized documents, acquire related evidence progressively, and verify fine-grained information against the source files.
    \item We evaluate DocFS on MMLongBench-Doc and FinRAGBench-V, achieving state-of-the-art performance on both benchmarks and outperforming the strongest baselines by 2.9 and 4.3 percentage points, respectively. Further analyses show that DocFS is up to 10.3$\times$ faster and consumes up to 10.1$\times$ fewer tokens than the strongest baseline during reasoning.
\end{itemize}

\section{Related Works}

\subsection{Retrieval-Based DocQA}

Retrieval-based DocQA methods retrieve query-relevant evidence from document representations to support answer generation. Existing methods can be divided into visual-representation-based retrieval and structure-aware retrieval, depending on how document information is represented and organized for evidence retrieval. The former primarily retrieves page images or visual representations: ColPali \cite{faysse2025colpali} uses multi-vector page representations, DSE \cite{ma2024unifying} adopts document screenshots for multimodal retrieval, and VisRAG \cite{yu2025visrag} retrieves and processes document images directly. M3DocRAG \cite{cho2024m3docrag} extends page retrieval to multimodal QA over multiple documents, while VDocRAG \cite{tanaka2025vdocrag} explores self-supervised pretraining for visually rich document retrieval. Although these methods preserve visual information, they do not explicitly organize document hierarchies and cross-modal relationships into structures that can be directly exploited. Structure-aware methods incorporate such information into retrieval: BookRAG \cite{wang2025bookrag} combines hierarchical content trees with entity-relation graphs, MHier-RAG \cite{gong2026mhier} retrieves chunks at multiple granularities, MoLoRAG \cite{wu2025molorag} traverses page graphs, and PDF-WuKong \cite{xie2026wukong} performs query-guided sampling of textual and visual blocks. Despite these advances, document structures primarily support evidence retrieval rather than continued exploration during answer reasoning. Consequently, the benefits of structured document representations remain closely tied to retrieval, rather than being fully exploited throughout the question-answering process.

\subsection{Agentic DocQA}

Agentic DocQA enables agents to acquire evidence and refine reasoning through iterative interactions with document-based environments. Existing methods differ in how they organize document content and expose the resulting representations for agent interaction. One group primarily supports agent interaction through iterative evidence retrieval, progressive evidence accumulation, or specialized agent coordination, with document access largely centered on pages, content units, or retrieved results. Doc-React \cite{wu2025doc} refines retrieval using information-gain and uncertainty feedback, SimpleDoc \cite{jain2025simpledoc} progressively accumulates retrieved evidence in working memory, and DocLens \cite{zhu2026doclens} narrows the search from documents to fine-grained visual elements. MACT \cite{yu2026visual} decomposes document analysis among specialized agents, while MDocAgent \cite{han2025mdocagent} coordinates modality-specific agents. These methods improve evidence acquisition and reasoning workflows, but do not necessarily provide a unified, explicitly organized document structure that agents can navigate to explore relationships across sections and modalities. Another group introduces structured representations to support agent interaction. DocAgent \cite{sun2025docagent} builds a tree-structured document outline with an interactive reading interface, DocAtlas \cite{wei2026docatlas} maintains a mutable document tree and note store, and MARDoc \cite{chen2026mardoc} retrieves over fine-grained semantic outlines. DocSeeker \cite{yan2026docseeker} grounds structured visual reasoning in evidence located in long documents. However, the relationship between document representations and agent access remains system-dependent: the operations available to agents are often designed around particular representations rather than provided through a common interface for exploring the organized document. Consequently, explicit document structures do not necessarily translate into directly accessible and consistently navigable environments for agents. DocFS addresses this gap by organizing document hierarchies and cross-modal relationships within a hierarchical file system and exposing the resulting content through standard file system operations, allowing agents to navigate, search, and read document information throughout question answering.

\section{Method}

\begin{figure*}[!t]
 \includegraphics[width=\textwidth]{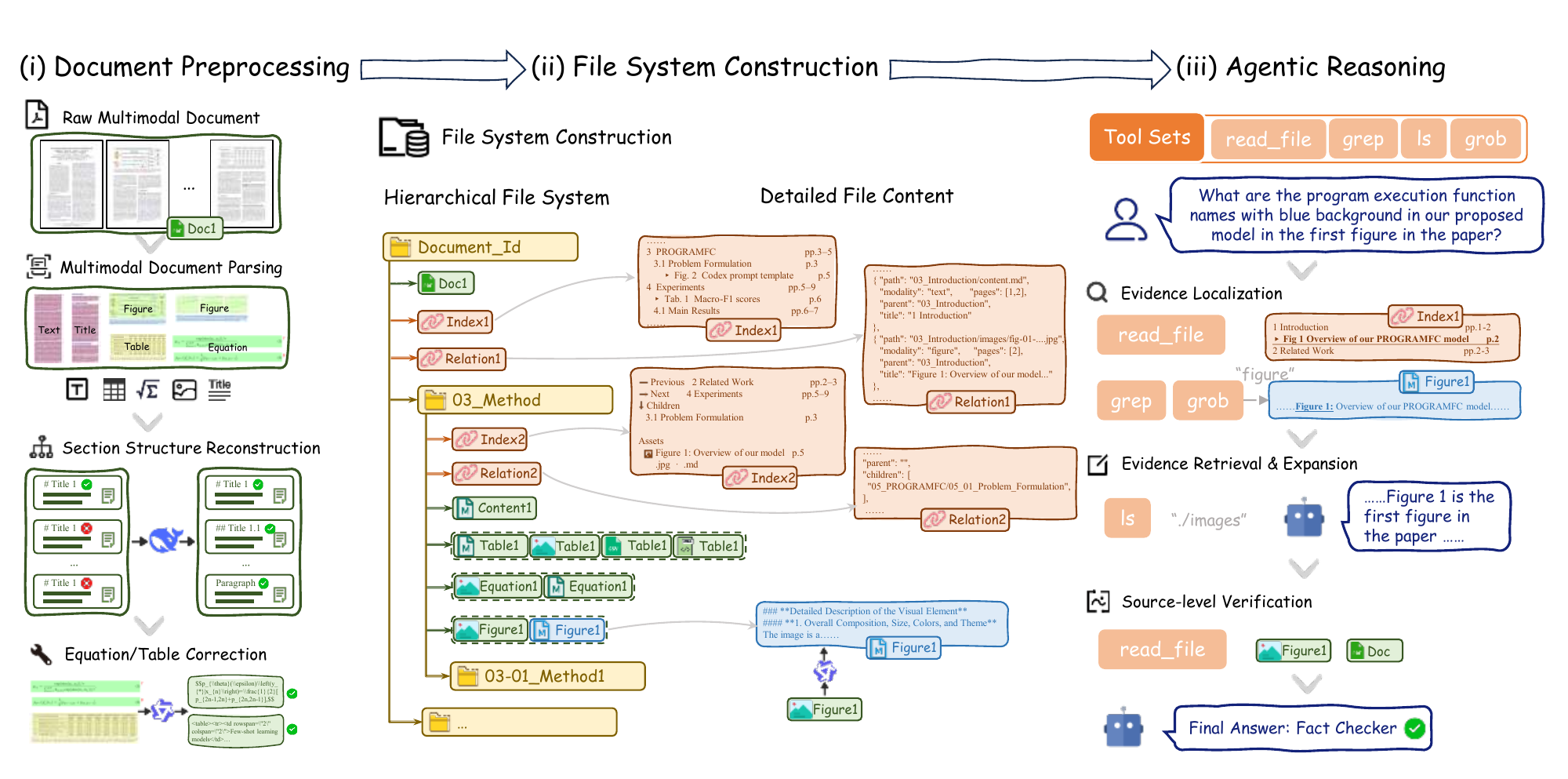}
  \caption{An overview of the agentic framework of DocFS.}
  \label{fig:method}
\end{figure*}

Figure~\ref{fig:method} illustrates an overview of DocFS, which consists of three stages. Document preprocessing (Section~\ref{sec:preprocessing}) parses the raw multimodal document and refines its structural hierarchy and content. File system construction (Section~\ref{sec:construction}) reorganizes the refined document into a hierarchical file system that preserves section structures and cross-modal associations, and exposes the organized content to the agent through a set of file system tools. Agentic reasoning (Section~\ref{sec:reasoning}) lets the agent explore the constructed file system with these tools, localizing candidate evidence and verifying it against the source files to produce reliable answers.

\subsection{Document Preprocessing}
\label{sec:preprocessing}

Before constructing the hierarchical file system, we parse the input multimodal documents and refine the resulting representations to obtain reliable document content and structural information for subsequent organization. This process consists of two steps: multimodal document parsing (Section~\ref{sec:parse}) and structural and content refinement (Section~\ref{sec:refine}).

\subsubsection{Multimodal Document Parsing}
\label{sec:parse}

We employ a document parsing tool to perform page-level parsing of the input documents, extracting multimodal elements, including \verb|Text|, \verb|Title|, \verb|Figure|, \verb|Table|, and \verb|Equation|, while preserving their content, type, page number, and bounding-box coordinates. This process converts each document into an initial representation comprising multimodal elements and their associated metadata, providing the basis for subsequent structural reconstruction and document organization.

\subsubsection{Structural and Content Refinement}
\label{sec:refine}

The initial parsing results may not fully preserve the hierarchical structure of the original document, and complex visual content may contain recognition errors. We therefore refine the parsing results along two dimensions: document structure and content accuracy.
\begin{itemize}
\item \textbf{Section Hierarchy Reconstruction}. We extract all elements identified as \verb|Title|, together with their contextual information, and feed them into a large language model to examine and reconstruct the hierarchical relationships among titles. This process addresses two primary issues. First, hierarchy flattening occurs when the parsing tool assigns the same hierarchical level to titles that originally belong to different levels. We correct these assignments to recover the appropriate section hierarchy, including cases where a title initially classified as a top-level heading should instead be assigned to a deeper level, such as Level 3. Second, title misclassification occurs when non-title elements are incorrectly identified as \verb|Title|. We identify such cases and reclassify the corresponding elements as body text. These corrections restore the hierarchical relationships between sections and subsections.
\item \textbf{Equation and Table Content Correction}. To address recognition errors in equations and tables, we leverage visual information from the original document pages for verification. Specifically, we extract screenshots of the corresponding equation and table regions and provide them to a vision-language model, which verifies and corrects the parsed LaTeX representations of equations and HTML representations of tables. This step improves the reliability of the parsed content, particularly for visually complex elements.
\end{itemize}

After these steps, we obtain a refined document representation containing multimodal elements, reconstructed section hierarchies, and corrected content, which serves as the foundation for subsequent hierarchical file system construction.

\subsection{File System Construction}
\label{sec:construction}

File System Construction is the core component of our approach. The resulting organization must not only capture the complex structure of multimodal documents, including hierarchical section structures and cross-modal associations, but also support effective access to and utilization of these structures by agents. To this end, we first reorganize preprocessed documents into a hierarchical directory structure, where nested directories mirror the section hierarchy and content is stored in type-specific files (Section~\ref{sec:fs-directory}). We then construct index files and relation files, which explicitly encode document structure and content relationships (Section~\ref{sec:fs-structure}). Finally, we generate semantic descriptions for images, enabling them to be retrieved through textual queries (Section~\ref{sec:fs-visual}). In addition, we retain the original document as the source PDF and preserve the correspondence between the organized content and its source pages, providing a direct reference for subsequent evidence verification. Through these designs, the file system preserves document hierarchies and cross-modal associations while transforming complex multimodal documents into an information space that agents can explore through a set of file system tools, which we describe in Section~\ref{sec:fs-tools}.

\subsubsection{Hierarchical Directory Construction}
\label{sec:fs-directory}

We reconstruct the preprocessed document representations into a hierarchical file system, organizing document content along two dimensions: directory hierarchy and file type. At the directory level, we construct nested directories according to the reconstructed section hierarchy, reflecting the hierarchical relationships between sections and subsections. The content of each section, together with its associated multimodal assets, is organized within the corresponding directory. This design explicitly maps the original document's section structure to an accessible directory hierarchy, allowing agents to explore document content according to its organizational structure rather than relying solely on a flattened representation. At the file level, we store different content elements in file formats appropriate to their data types. For example, textual content is stored in Markdown files, visual content (figures and tables) in image formats such as JPG, and tables also in CSV and HTML files. This design represents different modalities as independently accessible files in suitable formats, allowing agents to read and process them through corresponding file system tools. By combining directory hierarchy with file type, the file system simultaneously captures the section affiliation and data type of each content element, such that the document structure no longer resides merely in the metadata of the parsing results but is directly embodied in the accessible file organization.

\subsubsection{Structure Description Enhancement}
\label{sec:fs-structure}

Beyond its hierarchical section structure, a document contains numerous semantic relationships among its contents. To capture these relationships and facilitate access, we introduce two kinds of structure description files, index files and relation files, to support navigation and relationship-based information access, respectively. Index files record the locations of sections and associated content, providing navigation entry points through which agents can identify potentially relevant sections and content based on a query. Relation files explicitly record relationships between sections, between sections and multimodal assets, and among assets of different modalities. These relationships allow agents to discover and access related content after locating candidate evidence, thereby acquiring complementary information across sections and modalities. Specifically, we create index files and relation files at both the root directory and individual section directories, establishing a navigation and relationship mechanism that operates at both global and local levels. At the root level, the index file provides section-level navigation and global content lookup, while the relation file records relationships across the document. At the section level, the index file supports navigation within the current section, while the relation file captures relationships relevant to that section. This hierarchical design enables agents to locate potentially relevant content from a document-wide perspective and, upon entering a specific section, to further access related resources, without having to rely exclusively on global retrieval or traverse the directory tree level by level.

\subsubsection{Image Description Enhancement}
\label{sec:fs-visual}

Although the hierarchical file system can organize content of different modalities and establish content associations among its components, the semantic content of images remains difficult to retrieve based solely on file paths, file names, or section information. To enable images to participate in evidence-based localization in textual form, we further generate textual semantic descriptions for images, thereby providing keyword-searchable entry points to their content.

Specifically, we feed screenshots of images extracted from the document into a vision-language model to generate textual descriptions of their main content and relevant semantic characteristics. These descriptions are stored as searchable text files, with explicit mappings maintained to their corresponding original images. The generated descriptions provide a textual retrieval interface for images, allowing agents to identify potentially relevant visual evidence using keywords from a query. Through these mappings, agents can subsequently access the corresponding original images for further examination of their visual content.

Importantly, image semantic descriptions serve to assist visual evidence retrieval and localization rather than to replace the original visual content. For questions involving local objects, precise numerical values, or fine-grained visual details, agents must still examine the original images to avoid overlooking critical information that may not be captured in the textual descriptions.

\subsubsection{File System Tools}
\label{sec:fs-tools}

The agent interacts with the constructed file system through a set of file system tools, which provide the basic operations required to explore and use the organized document content. \verb|read_file| reads the content of a file. For text files, it reads the textual content. For figures, tables, equations, and other multimodal assets, it reads the corresponding file so that the agent can directly examine the visual content. For the source PDF, it accepts \verb|page_start| and \verb|page_end| to specify the page range to read, which supports source-level verification. \verb|grep| searches file contents by keyword and returns the matching content together with the corresponding file paths, which serves as the global retrieval operation for locating candidate evidence. \verb|ls| lists the entries of a directory, so that the agent can inspect which sections and files are available at the current level. \verb|glob| matches file paths by pattern, so that the agent can collect files of a given type or under a given directory in a single step.

These tools expose the file system as a set of standard operations, allowing the agent to navigate, retrieve and read information without task-specific interfaces.

\subsection{Agentic Reasoning}
\label{sec:reasoning}

Building on the constructed hierarchical file system, we design an agentic reasoning workflow in which an agent explores the organized document space through file system tools. The agent first identifies candidate evidence using the document's section hierarchy and global retrieval (Section~\ref{sec:loc}), then retrieves related context through relation files (Section~\ref{sec:expand}), and finally examines the source files---the original visual files and the source PDF---to verify fine-grained content (Section~\ref{sec:verify}).

\subsubsection{Evidence Localization}
\label{sec:loc}

The agent combines hierarchical navigation with global keyword search to locate question-relevant document content, progressively narrowing the search space to identify candidate evidence.

First, the agent reads the target document's directory structure and identifies potentially relevant sections based on the semantics of the question. The hierarchical directory structure reflects the document's section organization, enabling the agent to navigate the document hierarchy and explore potentially relevant content. The agent then retrieves globally using keywords from the question to identify relevant textual content, and obtains the corresponding file paths to form a set of candidate evidence.

By combining section-level navigation with global retrieval, the agent can identify promising search regions based on the document structure while directly locating candidate evidence distributed across different sections and modalities.

\subsubsection{Evidence Retrieval and Expansion}
\label{sec:expand}

After identifying candidate evidence, the agent reads the corresponding files and uses the explicitly encoded relationships in relation files to retrieve other relevant sections and multimodal content. This process supplements information distributed across different sections and modalities, allowing the retrieved evidence to be considered jointly in subsequent reasoning.

\subsubsection{Source-level Verification}
\label{sec:verify}

For visual evidence involving figures, tables, and equations, the agent directly accesses the corresponding original visual files, which are screenshots cropped from the source PDF, and understands their content. Information retrieved in previous stages is also used to facilitate interpretation.

Directly examining the original visual files lets the agent inspect fine-grained visual details and local information that may not be fully preserved in textual indexes or content descriptions. These details are then integrated with the textual evidence and related information retrieved in previous stages to support answer generation.

Moreover, the agent identifies the corresponding pages in the source PDF to verify the evidence supporting the answer, mitigating the impact of parsing errors or information loss on the final answer.

\section{Experiments}

To demonstrate the effectiveness of DocFS, we conduct comprehensive experiments and present the setup, main results, and analysis in this section. Specifically, we focus on the following research questions (RQ): \textbf{RQ1:} Does DocFS outperform existing state-of-the-art methods on complex multimodal document question answering? \textbf{RQ2:} How do the file system structure and the file system tools each contribute to the overall performance of DocFS? \textbf{RQ3:} How efficient is DocFS in terms of latency and token consumption? \textbf{RQ4:} How does the agent use the constructed file system in practice, in terms of tool usage, file access, and reasoning behavior?

\begin{table*}[htbp]
\centering
\small 
\caption{Main results on MMLongBench-Doc and FinRAGBench-V. Bold marks the best score in each column among all methods, and underline marks the second best. Methods marked with $^{\dagger}$ report results taken from DocLens, and those marked with $^{\ddagger}$ report results taken from their original papers; ``-'' indicates that the corresponding result is not reported in that source.}
\label{tab:main}
\setlength{\tabcolsep}{4pt} 
\renewcommand{\arraystretch}{1.1} 
\begin{tabular}{l|cccccccc|cccc} 
\toprule
\multirow{2}{*}{\textbf{Model}} & \multicolumn{8}{c|}{\textbf{MMLongBench-Doc}} & \multicolumn{4}{c}{\textbf{FinRAGBench-V}} \\
\cmidrule(lr){2-9} \cmidrule(lr){10-13}
 & \textbf{TXT} & \textbf{LAY} & \textbf{CHA} & \textbf{TAB} & \textbf{FIG} & \textbf{UNA} & \textbf{ACC} & \textbf{F1} & \textbf{TXT} & \textbf{TAB} & \textbf{CHA} & \textbf{ACC} \\
\midrule


\multicolumn{13}{c}{\textit{VLM Augmented with OCR}} \\
\midrule
Gemini-2.5-Pro$^{\dagger}$ & 59.7 & 65.3 & 60.8 & 68.3 & 55.7 & 58.4 & 63.3 & - & 70.0 & 70.0 & 56.2 & 64.9  \\
Qwen3.5-Plus & 64.7 & \textbf{71.3} & 58.7 & \textbf{74.9} & 64.8 & 63.7 & 66.6 & 65.4 & 80.2 & 78.3 & 74.2  & 77.2  \\
DeepSeek-V4.1-Flash & 62.9 & 66.6 & 59.5 & 69.0 & 60.8 & 59.8 & 63.3 & 62.3 & 80.6 & 75.9 & 70.9 & 75.0  \\
\midrule

\multicolumn{13}{c}{\textit{VLM-based RAG Frameworks}} \\
\midrule
M3DocRAG (w/ Qwen2-VL-7B)$^{\dagger}$ & 30.0 & 23.5 & 18.9 & 20.1 & 20.8 & 5.8 & 21.0 & 22.6 & - & - & - & - \\
BookRAG (w/ Qwen2.5VL-30B)$^{\ddagger}$ & - & - & - & - & - & - & 43.8 & 44.9 & - & - & - & - \\
MHier-RAG (w/ Qwen-turbo)$^{\ddagger}$ & 45.9 & 34.4 & 44.9 & 51.1 & 37.5 & \underline{76.2} & 52.3 & 46.0 & - & - & - & - \\
\midrule

\multicolumn{13}{c}{\textit{VLM-based Agentic Frameworks}} \\
\midrule
MACT (w/ MiMo-VL-7B)$^{\dagger}$ & - & - & - & - & - & - & 47.4 & - & - & - & - & - \\
MDocAgent (w/ GPT-4o)$^{\dagger}$ & - & - & - & - & - & - & 42.0 & - & - & - & - & - \\
SimpleDoc (w/ Gemini-2.5-Pro)$^{\dagger}$ & 48.4 & 54.8 & 55.7 & 56.1 & 52.5 & 59.7 & 56.6 & - & 67.5 & 64.0 & 60.9 & 63.6 \\
\midrule

\multicolumn{13}{!{\hspace*{-\tabcolsep}}l!{\hspace*{-\tabcolsep}}}{\textbf{DocLens}} \\
\quad w/ Qwen3.5-Plus & 63.7 & \underline{71.2} & 58.7 & 72.3 & \underline{65.0} & 75.4 & 68.7 & 66.1 & 85.7 & \textbf{83.3} & 68.8 & \underline{78.0} \\
\quad w/ DeepSeek-V4.1-Flash & 65.5 & 66.7 & 59.7 & 71.0 & 62.2 & 74.1 & 67.8 & 66.1 & 76.8 & 74.7 & 71.9 & 74.4 \\
\midrule

\multicolumn{13}{!{\hspace*{-\tabcolsep}}l!{\hspace*{-\tabcolsep}}}{\cellcolor{gray!20}\textbf{DocFS (Ours)}} \\
\quad w/ Qwen3.5-Plus & \textbf{70.6} & 68.4 & \textbf{68.0} & \underline{74.4} & \textbf{67.9} & 73.7 & \textbf{71.6} & \textbf{70.0} & \underline{89.1} & 79.9 & \textbf{80.9} & \textbf{82.3} \\
\quad w/ DeepSeek-V4.1-Flash & \underline{67.1} & 67.5 & \underline{64.1} & 71.3 & 63.3 & \textbf{76.3} & \underline{69.7} & \underline{67.5} & \textbf{92.7} & \underline{80.4} & \underline{78.2} & \textbf{82.3} \\
\bottomrule
\end{tabular}
\end{table*}

\subsection{Experimental Setup}

\paragraph{Benchmarks and Metrics}

We evaluate our approach on two challenging benchmarks: MMLongBench-Doc and FinRAGBench-V. MMLongBench-Doc comprises lengthy documents spanning multiple domains, requiring models to identify and integrate evidence distributed across different document regions and modalities. Its unanswerable questions also enable us to examine whether the proposed approach can avoid generating answers unsupported by the raw documents. FinRAGBench-V complements this evaluation with documents featuring dense and complex layouts. These characteristics allow us to further investigate the effectiveness of our file-system-based agentic workflow in supporting complex multimodal document question answering. We follow the original evaluation protocols established by the respective benchmarks, using rule-based scoring for MMLongBench-Doc and LLM-as-a-judge for FinRAGBench-V. 

\paragraph{Baselines and Models}

We evaluate our proposed agentic framework on two state-of-the-art models: Qwen3.5-Plus and DeepSeek-V4.1-Flash \cite{deepseekai2026deepseekv41flash}. Qwen3.5-Plus is evaluated with thinking mode enabled, whereas DeepSeek-V4.1-Flash operates in non-thinking mode. We compare our method with the following three baseline approaches:
\textbf{(1) VLM augmented with OCR:} Models that are directly fed with page screenshots and the text extracted by an OCR tool, without any external parsing or retrieval. We report the results of Gemini-2.5-Pro \cite{comanici2025gemini}, Qwen3.5-Plus and DeepSeek-V4.1-Flash;
\textbf{(2) VLM-based RAG Frameworks:} Methods that retrieve relevant content to augment the model input for generation. We compare with three representative RAG methods: M3DocRAG, BookRAG, and MHier-RAG;
\textbf{(3) VLM-based Agentic Frameworks:} Existing agentic methods, including MACT, MDocAgent, SimpleDoc, and DocLens. 
For DocLens, the most recent and best-performing training-free framework, we reproduce its results with both backbone models to ensure a fair comparison. For the other methods, we report the best scores reported in either their original papers or DocLens.

Further statistics and implementation details are provided in Appendix~\ref{app:exp_details}.

\subsection{Main Results - RQ1}

Table~\ref{tab:main} presents the experimental results of DocFS on two benchmarks, MMLongBench-Doc and FinRAGBench-V. \textbf{Overall, DocFS achieves the highest accuracy, outperforming DocLens, the strongest training-free baseline, across both benchmarks and backbone models.} On MMLongBench-Doc, DocFS reaches 71.6\% accuracy and 70.0\% F1 scores with Qwen3.5-Plus, gains of 2.9 and 3.9 points over DocLens, and 69.7\% and 67.5\% with DeepSeek-V4.1-Flash, gains of 1.9 and 1.4 points. On FinRAGBench-V, both backbones reach 82.3\% accuracy, corresponding to gains of 4.3 and 7.9 points, respectively. These results demonstrate the effectiveness of organizing complex multimodal documents into a file system and enabling agents to perform question answering through the resulting structure.

\textbf{Strengths: Effective Utilization of Textual and Visual Evidence.} A closer examination of performance across evidence types reveals that DocFS performs competitively on questions involving textual content and various visual elements. On MMLongBench-Doc with Qwen3.5-Plus, the largest improvements are on chart (+9.3) and text (+6.9) questions, with smaller but consistent gains on figure (+2.9) and table (+2.1) questions; DeepSeek-V4.1-Flash preserves the same ordering (chart +4.4, text +1.6, figure +1.1, table +0.3). FinRAGBench-V exhibits the same tendency more strongly, with chart accuracy rising by 12.1 and 6.3 points for the two backbones and text rising by 15.9 points with DeepSeek-V4.1-Flash. These results indicate that the benefits of DocFS extend beyond text retrieval to evidence involving different visual modalities. By organizing document content according to its structure and using element-type files and structure description files to facilitate agent access to relevant evidence, DocFS provides a unified organizational foundation for locating and utilizing information across modalities.

\subsection{Ablation Study - RQ2}
\label{sec:ablation}

\begin{table}[t]
\centering
\footnotesize
\setlength{\tabcolsep}{4pt}
\renewcommand{\arraystretch}{1.15}
\caption{Ablation study on MMLongBench-Doc and FinRAGBench-V.}
\label{tab:ablation}
\begin{threeparttable}
\begin{tabular}{l ccc}
\toprule
\multirow{2}{*}{\textbf{Variants}} & \multicolumn{2}{c}{\textbf{MMLongBench-Doc}} & \textbf{FinRAGBench-V} \\
\cmidrule(lr){2-3} \cmidrule(lr){4-4}
 & \textbf{ACC} & \textbf{F1} & \textbf{ACC} \\
\midrule
\textbf{DocFS (Qwen3.5-Plus)} & \textbf{71.6} & \textbf{70.0} & \textbf{82.3} \\
\rowcolor{gray!15}\multicolumn{4}{@{}l}{(a) File System Structure} \\
w/o File System & 66.6 & 64.7 & 78.8 \\
w/o Preprocessing & 66.7 & 63.9 & 80.6 \\
w/o Structure Desc. & 66.3 & 64.0 & 79.8 \\
w/o Image Desc. & 66.9 & 64.4 & 79.8 \\
\rowcolor{gray!15}\multicolumn{4}{@{}l}{(b) File System Tools} \\
w/o \verb|grep| & 62.0 & 59.3 & 74.0 \\
w/o \verb|ls| & 66.6 & 64.7 & 80.9 \\
\midrule
\textbf{DocFS (DeepSeek-V4.1-Flash)} & \textbf{69.7} & \textbf{67.5} & \textbf{82.3} \\
\rowcolor{gray!15}\multicolumn{4}{@{}l}{(a) File System Structure} \\
w/o File System & 64.4 & 63.3 & 71.9 \\
w/o Preprocessing & 58.7 & 57.2 & 81.6 \\
w/o Structure Desc. & 59.3 & 58.1 & 79.8 \\
w/o Image Desc. & 58.7 & 57.8 & 81.9 \\
\rowcolor{gray!15}\multicolumn{4}{@{}l}{(b) File System Tools} \\
w/o \verb|grep| & 56.6 & 55.6 & 79.8 \\
w/o \verb|ls| & 58.9 & 57.9 & 81.8 \\
\bottomrule
\end{tabular}
\end{threeparttable}
\end{table}

To investigate the effects of different components of DocFS on complex multimodal document question answering, we conduct a series of ablation experiments on two groups of variants that remove either a component of the file system structure or a tool provided by the file system.
\textbf{(1) w/o File System} uses the document parsing results directly to answer questions, without constructing a file system or applying any additional preprocessing or organizational mechanisms. 
\textbf{(2) w/o Preprocessing} constructs the file system directly from the raw parsing results, without applying any additional preprocessing, including section hierarchy reconstruction and equation and table content correction. 
\textbf{(3) w/o Structure Desc.} retains the document content and the basic directory structure of the file system, but removes the structure description files, i.e., all index and relation files, that facilitate document navigation and information retrieval. 
\textbf{(4) w/o Image Desc.} retains the image files and other organizational components of the file system but removes the textual descriptions generated for the images. 
\textbf{(5) w/o grep} removes the keyword search tool.
\textbf{(6) w/o ls} removes the directory listing tool.

Table~\ref{tab:ablation} presents the ablation results of DocFS using two backbone models on two benchmark datasets. Overall, the full DocFS achieves the highest overall accuracy across both backbones and both datasets, demonstrating the effectiveness of the proposed agentic framework in improving complex multimodal document question answering. Removing any component or tool lowers performance. Appendix~\ref{app:ablation} provides comprehensive per-variant analysis.

\textbf{The constructed file system and its structure components jointly contribute to the overall performance.} These results demonstrate that organizing parsed documents into a hierarchical file system, together with structure and image description files, yields consistent overall benefits across different backbone models and datasets. Removing preprocessing which contains section hierarchy reconstruction and equation and table content correction leads to a decrease in the overall accuracy on MMLongBench-Doc by 4.9 points for Qwen3.5-Plus and 11.0 points for DeepSeek-V4.1-Flash. Structure description files help agents effectively exploit organized structure: removing the index and relation files leads to a decrease in the overall accuracy for both backbone models on both datasets. Image descriptions improve the retrieval and utilization of visual evidence: the degradation is concentrated on questions whose evidence is visual. For Qwen3.5-Plus, the CHA and FIG subsets drop by 11.3 and 7.3 points, and for DeepSeek-V4.1-Flash, the CHA and FIG subsets drop by 15.8 and 10.8 points. These findings align with our original motivation for introducing image descriptions: converting key information in visual elements into textual descriptions enables agents to discover potentially relevant images through text-based retrieval and then inspect the original images for fine-grained understanding.

\textbf{Global keyword search and directory listing are both critical access paths.} Removing \verb|grep| causes the largest drop among all variants on MMLongBench-Doc. Without global search, the agent has to descend the directory tree level by level and guess which files contain the answer, so evidence scattered across distant sections is easily missed. Removing \verb|ls| also lowers the overall accuracy for both backbone models on both benchmarks. Without \verb|ls|, the agent can no longer inspect which sections and files exist at each level of the hierarchy, and has to reach candidate evidence only through structure description files and keyword search. The decline indicates that browsing the directory structure is a useful complement to content-based search: keyword retrieval locates evidence by matching query terms, while directory listing reveals the available search space itself. Removing either one weakens the agent's ability to exploit the organized structure.

\subsection{Efficiency Analysis - RQ3}

\begin{table}[t]
\centering
\caption{Efficiency comparison between DocLens and DocFS. \textbf{Abbreviations:} Qwen-Plus = Qwen3.5-Plus, DS-Flash = DeepSeek-V4.1-Flash.}
\label{tab:efficiency}
\begin{tabular}{llrrrr}
\toprule
\multirow{2}{*}{Backbone} & \multirow{2}{*}{Method}
  & \multicolumn{2}{c}{Time} & \multicolumn{2}{c}{Tokens} \\
\cmidrule(lr){3-4} \cmidrule(lr){5-6}
 & & (s) & Gain ($\times$) & (K) & Gain ($\times$) \\
\midrule
\multicolumn{6}{>{\columncolor{gray!20}}l}{\textit{MMLongBench-Doc}} \\
\multirow{2}{*}{Qwen-Plus}
   & DocLens & 359.9 & -- & 217.9 & -- \\
 & DocFS & \textbf{43.9} & \textbf{8.2} & \textbf{142.9} & \textbf{1.5} \\
\cmidrule{1-6}
\multirow{2}{*}{DS-Flash}
   & DocLens & 76.8 & -- & 407.0 & -- \\
 & DocFS & \textbf{17.8} & \textbf{4.3} & \textbf{180.9} & \textbf{2.3} \\
\midrule
\multicolumn{6}{>{\columncolor{gray!20}}l}{\textit{FinRAGBench-V}} \\
\multirow{2}{*}{Qwen-Plus}
   & DocLens & 436.1 & -- & 602.4 & -- \\
 & DocFS & \textbf{52.7} & \textbf{8.3} & \textbf{185.0} & \textbf{3.3} \\
\cmidrule{1-6}
\multirow{2}{*}{DS-Flash}
   & DocLens & 189.9 & -- & 1,555.9 & -- \\
 & DocFS & \textbf{18.4} & \textbf{10.3} & \textbf{154.7} & \textbf{10.1} \\
\bottomrule
\end{tabular}
\end{table}


\textbf{DocFS is substantially more efficient than the strongest baseline.}
Table~\ref{tab:efficiency} compares the reasoning latency and the total token consumption of DocFS and DocLens---the strongest training-free baseline, whose cost is the sum of its two phases. In all four settings, DocFS attains higher accuracy (Table~\ref{tab:main}) at a fraction of the cost. With Qwen3.5-Plus, DocFS reduces the mean latency from 359.9s to 43.9s on MMLongBench-Doc (8.2$\times$) and from 436.1s to 52.7s on FinRAGBench-V (8.3$\times$), and reduces the mean token consumption from 217.9K to 142.9K (1.5$\times$) and from 602.4K to 185.0K (3.3$\times$). With DeepSeek-V4.1-Flash, the mean latency drops from 76.8s to 17.8s (4.3$\times$) and from 189.9s to 18.4s (10.3$\times$), while the mean token consumption drops from 407.0K to 180.9K (2.3$\times$) and from 1,555.9K to 154.7K (10.1$\times$). Two observations follow. First, the cost of DocFS barely changes across the two benchmarks: its mean token consumption is 142.9K on MMLongBench-Doc versus 185.0K on FinRAGBench-V with Qwen3.5-Plus, and 180.9K versus 154.7K with DeepSeek-V4.1-Flash, while its mean latency moves by at most 1.2$\times$. By contrast, the token cost of DocLens grows sharply with document density: its mean token consumption is 2.8$\times$ higher on FinRAGBench-V than on MMLongBench-Doc with Qwen3.5-Plus (602.4K versus 217.9K) and 3.8$\times$ higher with DeepSeek-V4.1-Flash (1,555.9K versus 407.0K), whereas the token cost of DocFS changes by only 1.3$\times$ and 0.9$\times$, respectively. Second, the gain therefore widens where documents are denser and more layout-heavy: on FinRAGBench-V, DocLens has to process large amounts of page-level content in both phases, whereas DocFS follows the section hierarchy directly to the relevant files. These results confirm that exposing document structure through a file system, together with tools aligned with it, not only improves answer quality but also makes agentic reasoning substantially cheaper and less sensitive to document density.

All comparisons in Table~\ref{tab:efficiency} are measured on the reasoning stage. The document processing stage of both methods (pdf-to-images and parsing for DocLens and preprocessing and file system construction for DocFS) is executed offline and dominated by the concurrency of the underlying parsing and processing tools, so its wall-clock time is not directly comparable; we therefore restrict the time comparison to the reasoning stage. DocFS additionally incurs a document processing token cost that is not charged to it in the table; even when this cost is added, DocFS still consumes fewer tokens
than DocLens in all four settings (1.4--5.6$\times$), which is analyzed in Appendix~\ref{app:doc_processing_costs}.

\subsection{Detailed Analysis - RQ4}
\label{sec:detailed_analysis}

\textbf{Tool call distribution.}
Figure~\ref{fig:tool_call_distribution} in Appendix~\ref{app:detailed_analysis_figures} counts how often each of the four file system tools is called on both benchmarks, separately for the two backbones. \verb|read_file| dominates in all four settings, \verb|grep| ranks second, and \verb|glob| is almost never used, while \verb|ls| is used much more on FinRAGBench-V. This indicates that the agent mainly alternates between keyword search and file reading, and that the four generic tools are sufficient to support the whole workflow.

\textbf{File access distribution.}
Figure~\ref{fig:read_file_groups} further divides the \verb|read_file| calls into six file groups. Structure description files and textual content files together take about half of all reads, other files such as the source PDF take about one fifth, and the remaining reads are among table, visual, and image-description files. Table reads are clearly more frequent on FinRAGBench-V, whereas visual file reads are more frequent on MMLongBench-Doc, and image description files are the least read group. The agent therefore uses the navigation layer throughout, verifies evidence against the source files regularly, and reads the evidence type that is actually common in each benchmark.

\textbf{End-to-end reasoning traces.}
To complement the aggregate view above, Figure~\ref{fig:case_study} traces two representative queries end-to-end on MMLongBench-Doc (a) and FinRAGBench-V (b). Both follow the same three-phase pattern --- \emph{Localization} via \verb|index.md| and \verb|grep|, \emph{Retrieval} of the candidate table or figure, and \emph{Source Verification} against the source PDF --- but they stress different capabilities. Case (a) requires cross-modal grounding between a color-coded architecture figure (Section~9) and a numerical table (Section~10); the figure's VLM output directly reveals that ``blue'' maps to \emph{Conscientiousness}, so the visual-understanding step is folded into the read response. Case (b) shows the agent recovering from a dead end: the first \verb|grep| fails, the balance-sheet table lacks the 2019 column, and a follow-up \verb|grep("2019")| redirects it to the Stockholders' Equity table, which is then verified against PDF pages 75 and 77. These traces make concrete what the aggregate distributions abstract away: the file system lets the agent re-orient mid-trace and verify evidence at source level without rerunning any parser.

\section{Conclusion}

In this paper, we proposed DocFS, a file-system-based agentic framework for complex multimodal document question answering. DocFS reorganized raw documents into hierarchical file systems, where directories mirrored the section hierarchy and files held the textual and visual elements, together with structure and image description files that supported navigation and retrieval. On top of this environment, agents answered questions through standard file system tools, following a workflow of evidence localization, evidence retrieval and expansion, and source-level verification. Experiments on MMLongBench-Doc and FinRAGBench-V showed that paired with Qwen3.5-Plus, DocFS achieved state-of-the-art performance, outperforming the strongest training-free baseline by 2.9 and 4.3 points in overall accuracy on the two benchmarks. Ablation results showed that every component of the file system contributed to the final performance. Efficiency analysis further showed that DocFS was 4.3--10.3$\times$ faster and consumed 1.5--10.1$\times$ fewer tokens than the strongest baseline during reasoning. Despite these promising results, several challenges remain open. Future work could explore more adaptive document organization strategies for diverse document types and question requirements, develop richer interaction mechanisms for navigating complex cross-modal relationships, and investigate how agents can better balance evidence exploration costs against answer reliability. These directions could further advance the design of structured, actionable environments for agent-based reasoning over complex multimodal documents.




\bibliographystyle{ACM-Reference-Format}
\bibliography{reference}

\appendix

\begin{figure*}[t]
  \centering
  \includegraphics[width=\textwidth]{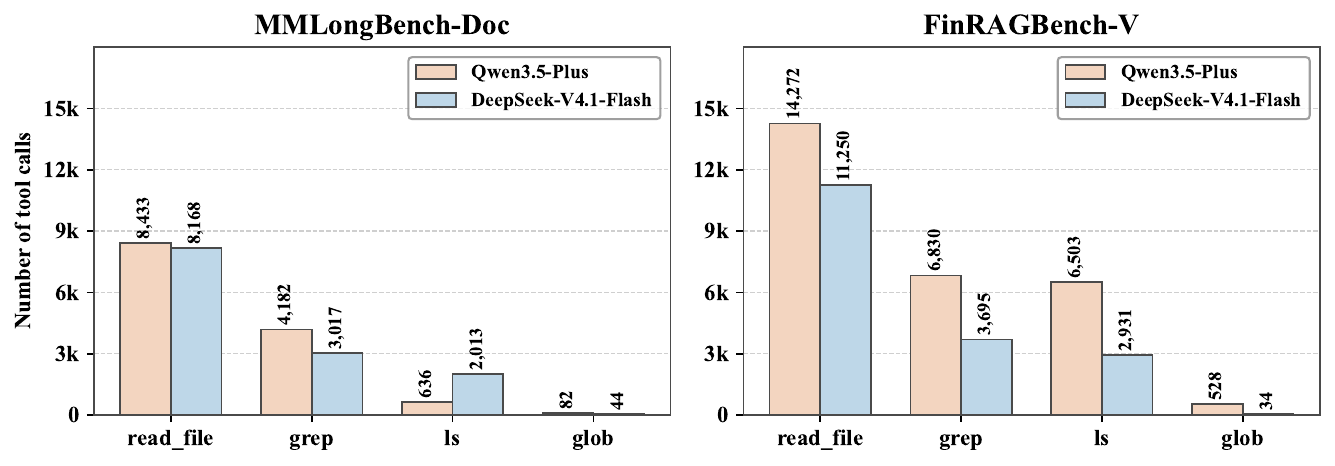}
  \caption{Number of tool calls per tool type for Qwen3.5-Plus and DeepSeek-V4.1-Flash on MMLongBench-Doc (left) and FinRAGBench-V (right).}
  \label{fig:tool_call_distribution}
\end{figure*}

\begin{figure*}[t]
  \centering
  \includegraphics[width=\textwidth]{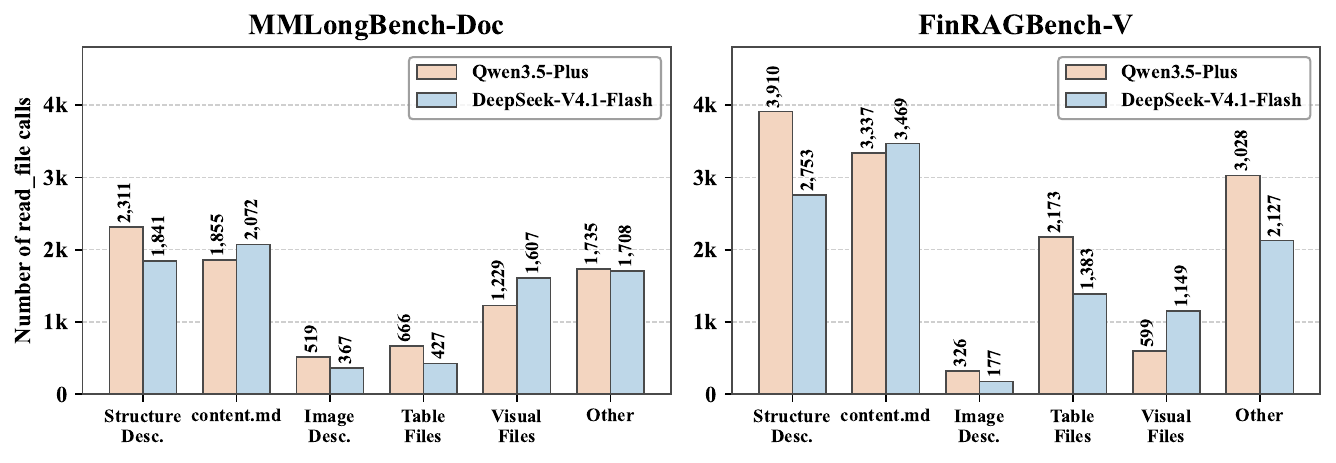}
  \caption{Distribution of successful \texttt{read\_file} calls across six file groups on MMLongBench-Doc (left) and FinRAGBench-V (right). Structure Desc. denotes index and relation files; content.md denotes the textual content of the document; Image Desc. denotes the generated textual descriptions of images; Table Files include CSV/HTML representations of tables; Visual Files are the original visual files such as figures and tables; Other denotes other files such as the source PDF.}
  \label{fig:read_file_groups}
\end{figure*}

\begin{figure*}[t]
\centering
\small
\setlength{\tabcolsep}{4pt}
\newcommand{\tbr}{\textcolor{gray!50}{|\_\_}\ }
\newcommand{\tind}[1]{\hspace*{#1em}}
\newcommand{\tdir}{\textcolor{blue!60!black}{\faFolder}\ }
\newcommand{\ttxt}{\textcolor{black!60}{\faFile[regular]}\ }
\newcommand{\timg}{\textcolor{green!50!black}{\faFileImage}\ }
\newcommand{\tcsv}{\textcolor{orange!80!black}{\faTable}\ }
\newcommand{\tpdf}{\textcolor{red!70!black}{\faFilePdf}\ }
\newcommand{\tgen}{\textcolor{black!60}{\faFile}\ }
\newcommand{\tvis}{~\textcolor{red!70!black}{\textbf{[*]}}}
\renewcommand{\arraystretch}{1.15}

\noindent\textbf{(a) MMLongBench-Doc} \\[2pt]
\begin{tabular}{>{\raggedright\arraybackslash}p{0.30\textwidth} >{\raggedright\arraybackslash}p{0.65\textwidth}}
\toprule
\textbf{Filesystem view} & \textbf{Agent reasoning trace} \\
\midrule
\begin{minipage}[t]{0.30\textwidth}
\scriptsize\ttfamily
\tbr\tdir dataspace/2303.05039v2/\\
\tind{1}\tbr\ttxt index.md\tvis\\
\tind{1}\tbr\tgen relation.json\\
\tind{1}\tbr\tpdf source.pdf\tvis\\
\tind{1}\tbr\tdir 09\_METHODS/\\
\tind{2}\tbr\tdir 09\_02\_Pers-enh\_NCF/\\
\tind{3}\tbr\ttxt content.md\\
\tind{3}\tbr\tdir images/\\
\tind{4}\tbr\timg fig-01-arch.jpg\tvis\\
\tind{4}\tbr\ttxt fig-01-arch.md\\
\tind{1}\tbr\tdir 10\_EXPERIMENTS/\\
\tind{2}\tbr\tdir 10\_02\_dist\_users/\\
\tind{3}\tbr\ttxt content.md\\
\tind{3}\tbr\tdir tables/\\
\tind{4}\tbr\tcsv tab-04-extr.csv\tvis\\
\tind{3}\tbr\tdir images/\\
\tind{4}\tbr\timg fig-02-dist.jpg\tvis\\
\tind{4}\tbr\ttxt fig-02-dist.md\\
\tind{1}\tbr\tdir 11\_DISCUSSION/\\
\end{minipage}
&
\begin{minipage}[t]{0.65\textwidth}
\scriptsize
\textbf{Q:} What are the personality scores of the data sample with extreme personality cases, in which the personality vector color is blue in the overall model structure? (answer in ascending list)\\
\textcolor{orange!80!black}{\textbf{[Phase 1] Evidence Localization}}\\
\textcolor{green!50!black}{\textbf{TOOL CALL}}\ \texttt{\textbf{\textcolor{purple!70!black}{read\_file}}\textbf{\textcolor{brown!60!black}{(index.md)}}} \\
\textcolor{gray}{\textbf{RESPONSE}} \texttt{[sec tree] ... 09\_02\_Pers-enh\_NCF/images/fig-01-arch.jpg [p.5]; 10\_02\_dist\_users/tables/tab-04-extr.csv [p.6] ...}\\
\textcolor{green!50!black}{\textbf{TOOL CALL}}\ \texttt{\textbf{\textcolor{purple!70!black}{grep}}\textbf{\textcolor{brown!60!black}{("extreme", dataspace/2303.05039v2/)}}} \\
\textcolor{gray}{\textbf{RESPONSE}} \texttt{10\_02\_dist\_users/tables/tab-04-extr.csv}\\
\textcolor{orange!80!black}{\textbf{[Phase 2] Evidence Retrieval \& Expansion}}\\
\textcolor{green!50!black}{\textbf{TOOL CALL}}\ \texttt{\textbf{\textcolor{purple!70!black}{read\_file}}\textbf{\textcolor{brown!60!black}{(tab-04-extr.csv)}}} \\
\textcolor{gray}{\textbf{RESPONSE}} \texttt{Openness 63.07, 62.62; Conscientiousness 75.38, 71.02; Extroversion 75.06, 72.90; ...}\\
\textcolor{green!50!black}{\textbf{TOOL CALL}}\ \texttt{\textbf{\textcolor{purple!70!black}{read\_file}}\textbf{\textcolor{brown!60!black}{(fig-01-arch.jpg)}}} \\
\textcolor{gray}{\textbf{RESPONSE}} \texttt{[VLM] Figure 1: Conscientiousness embedding drawn in \textcolor{blue}{\textbf{blue}} in both Method 1 and Method 2}\\
\textcolor{green!50!black}{\textbf{TOOL CALL}}\ \texttt{\textbf{\textcolor{purple!70!black}{read\_file}}\textbf{\textcolor{brown!60!black}{(fig-02-dist.jpg)}}} \\
\textcolor{gray}{\textbf{RESPONSE}} \texttt{[VLM] Figure 2: personality trait distribution chart}\\
\textcolor{orange!80!black}{\textbf{[Phase 3] Source-level Verification}}\\
\textcolor{green!50!black}{\textbf{TOOL CALL}}\ \texttt{\textbf{\textcolor{purple!70!black}{read\_file}}\textbf{\textcolor{brown!60!black}{(source.pdf, page\_start=6, page\_end=6)}}} \\
\textcolor{gray}{\textbf{RESPONSE}} \texttt{[PDF p.6] Table 4 content matches the parsed CSV}\\
\textcolor{green!50!black}{\textbf{TOOL CALL}}\ \texttt{\textbf{\textcolor{purple!70!black}{read\_file}}\textbf{\textcolor{brown!60!black}{(source.pdf, page\_start=5, page\_end=5)}}} \\
\textcolor{gray}{\textbf{RESPONSE}} \texttt{[PDF p.5] Figure 1 confirms Conscientiousness vector is blue}\\
\textcolor{blue!60!black}{\textbf{ANSWER}} \texttt{["71.02", "75.38"]}
\end{minipage}
\\
\bottomrule
\end{tabular}

\vspace{6pt}

\noindent\textbf{(b) FinRAGBench-V} \\[2pt]
\begin{tabular}{>{\raggedright\arraybackslash}p{0.30\textwidth} >{\raggedright\arraybackslash}p{0.65\textwidth}}
\toprule
\textbf{Filesystem view} & \textbf{Agent reasoning trace} \\
\midrule
\begin{minipage}[t]{0.30\textwidth}
\scriptsize\ttfamily
\tbr\tdir dataspace/NASDAQ\_AMBC\_2021/\\
\tind{1}\tbr\ttxt index.md\tvis\\
\tind{1}\tbr\tgen relation.json\\
\tind{1}\tbr\tpdf source.pdf\tvis\\
\tind{1}\tbr\tdir 07\_Item\_7\_MD\&A/\\
\tind{2}\tbr\tdir 07\_48\_BALANCE\_SHEET/\\
\tind{3}\tbr\ttxt content.md\tvis\\
\tind{1}\tbr\tdir 09\_Report\_of\_Indep\_Pub\_Ac/\\
\tind{2}\tbr\tdir 09\_06\_AMBAC\_SUB/\\
\tind{3}\tbr\ttxt content.md\\
\tind{3}\tbr\tdir tables/\\
\tind{4}\tbr\tcsv tab-00.csv\tvis\\
\tind{2}\tbr\tdir 09\_07\_AMBAC\_SUB/\\
\tind{3}\tbr\tdir tables/\\
\tind{4}\tbr\tcsv tab-00.csv\tvis\\
\tind{3}\tbr\tgen relation.json\tvis\\
\tind{3}\tbr\ttxt index.md\\
\end{minipage}
&
\begin{minipage}[t]{0.65\textwidth}
\scriptsize
\textbf{Q:} According to Ambac Financial Group, Inc. 2021 FORM 10-K, compare the total balance between the fiscal years 2019 and 2020.\\
\textcolor{orange!80!black}{\textbf{[Phase 1] Evidence Localization}}\\
\textcolor{green!50!black}{\textbf{TOOL CALL}}\ \texttt{\textbf{\textcolor{purple!70!black}{ls}}\textbf{\textcolor{brown!60!black}{(dataspace/NASDAQ\_AMBC\_2021/)}}} \\
\textcolor{gray}{\textbf{RESPONSE}} \texttt{34 sections: 06\_Item\_6, 07\_Item\_7\_MD\&A, 09\_Report\_of\_Indep\_Pub\_Ac, ...}\\
\textcolor{green!50!black}{\textbf{TOOL CALL}}\ \texttt{\textbf{\textcolor{purple!70!black}{read\_file}}\textbf{\textcolor{brown!60!black}{(index.md)}}} \\
\textcolor{gray}{\textbf{RESPONSE}} \texttt{[TOC] sec 09 = Financial Statements; 07\_48 = BALANCE\_SHEET\_in\_millions}\\
\textcolor{green!50!black}{\textbf{TOOL CALL}}\ \texttt{\textbf{\textcolor{purple!70!black}{grep}}\textbf{\textcolor{brown!60!black}{("CONSOLIDATED BALANCE SHEET", ..., files\_with\_matches)}}} \\
\textcolor{gray}{\textbf{RESPONSE}} \texttt{09\_06\_AMBAC\_SUB}\\
\textcolor{orange!80!black}{\textbf{[Phase 2] Evidence Retrieval \& Expansion}}\\
\textcolor{green!50!black}{\textbf{TOOL CALL}}\ \texttt{\textbf{\textcolor{purple!70!black}{read\_file}}\textbf{\textcolor{brown!60!black}{(07\_48\_BALANCE\_SHEET/content.md)}}} \\
\textcolor{gray}{\textbf{RESPONSE}} \texttt{MD\&A narrative --- no 2019 vs 2020 comparison table}\\
\textcolor{green!50!black}{\textbf{TOOL CALL}}\ \texttt{\textbf{\textcolor{purple!70!black}{read\_file}}\textbf{\textcolor{brown!60!black}{(09\_06\_AMBAC\_SUB/tables/tab-00.csv)}}} \\
\textcolor{gray}{\textbf{RESPONSE}} \texttt{Consolidated Balance Sheet --- columns are Dec 31, 2021 / Dec 31, 2020 only (no 2019)}\\
\textcolor{green!50!black}{\textbf{TOOL CALL}}\ \texttt{\textbf{\textcolor{purple!70!black}{grep}}\textbf{\textcolor{brown!60!black}{("2019", ..., files\_with\_matches)}}} \\
\textcolor{gray}{\textbf{RESPONSE}} \texttt{09\_07\_AMBAC\_SUB (Consolidated Statements of Stockholders' Equity)}\\
\textcolor{green!50!black}{\textbf{TOOL CALL}}\ \texttt{\textbf{\textcolor{purple!70!black}{read\_file}}\textbf{\textcolor{brown!60!black}{(09\_07\_AMBAC\_SUB/tables/tab-00.csv)}}} \\
\textcolor{gray}{\textbf{RESPONSE}} \texttt{Balance at Dec 31, 2019: \$1,536; Balance at Dec 31, 2020: \$1,140 (Total stockholders' equity, \$M)}\\
\textcolor{green!50!black}{\textbf{TOOL CALL}}\ \texttt{\textbf{\textcolor{purple!70!black}{read\_file}}\textbf{\textcolor{brown!60!black}{(09\_07\_AMBAC\_SUB/relation.json)}}} \\
\textcolor{gray}{\textbf{RESPONSE}} \texttt{source\_page: [77] (PDF p.77, 10-K p.67)}\\
\textcolor{orange!80!black}{\textbf{[Phase 3] Source-level Verification}}\\
\textcolor{green!50!black}{\textbf{TOOL CALL}}\ \texttt{\textbf{\textcolor{purple!70!black}{read\_file}}\textbf{\textcolor{brown!60!black}{(source.pdf, page\_start=77, page\_end=77)}}} \\
\textcolor{gray}{\textbf{RESPONSE}} \texttt{[PDF p.77] Stockholders' Equity table matches the parsed CSV}\\
\textcolor{green!50!black}{\textbf{TOOL CALL}}\ \texttt{\textbf{\textcolor{purple!70!black}{read\_file}}\textbf{\textcolor{brown!60!black}{(source.pdf, page\_start=75, page\_end=76)}}} \\
\textcolor{gray}{\textbf{RESPONSE}} \texttt{[PDF p.75] Consolidated Balance Sheets confirm total stockholders' equity = \$1,140 at Dec 31, 2020}\\
\textcolor{blue!60!black}{\textbf{ANSWER}} \texttt{Dec 31, 2019: \$1,536M; Dec 31, 2020: \$1,140M; decreased \$396M}
\end{minipage}
\\
\bottomrule
\end{tabular}

\caption{End-to-end reasoning traces of DocFS on one MMLongBench-Doc query (a) and one FinRAGBench-V query (b). For each case, the left column shows the organized filesystem with paths actually visited marked by~\textbf{\textcolor{red!70!black}{[*]}}; the right column groups filesystem tool calls (\textcolor{green!50!black}{\textbf{TOOL CALL}}) and observed results (\textcolor{gray}{\textbf{RESPONSE}}) by phase. All operations are standard file system tools (\texttt{ls}, \texttt{read\_file}, \texttt{grep}).}
\label{fig:case_study}
\end{figure*}

\section{Experimental Details}
\label{app:exp_details}

\subsection{Dataset Details}

We evaluate DocFS on two benchmarks that target complex multimodal document question answering, MMLongBench-Doc and FinRAGBench-V. MMLongBench-Doc covers documents from seven domains, including research reports, tutorials, academic papers, guidebooks, brochures, administration and industry documents, and financial reports, and contains both answerable and unanswerable questions. FinRAGBench-V focuses on financial reports with dense and complex layouts. The full statistics of the two benchmarks are summarized in Table~\ref{tab:dataset_stats}.

\begin{table}[h]
\centering
\footnotesize
\setlength{\tabcolsep}{4pt}
\renewcommand{\arraystretch}{1.15}
\caption{Dataset statistics of MMLongBench-Doc and FinRAGBench-V.}
\label{tab:dataset_stats}
\begin{tabular}{l cc}
\toprule
\multirow{2}{*}{\textbf{Statistics}} & \multicolumn{2}{c}{\textbf{Dataset}} \\
\cmidrule(lr){2-3}
 & \textbf{MMLongBench-Doc} & \textbf{FinRAGBench-V} \\
\midrule
\textbf{Documents} & 135 & 301 \\
\quad Avg./Median pages & 47.5 / 28 & 76.1 / 57.0 \\
\quad Avg./Median words & 8,393 / 5,743 & 36,026 / 16,329 \\
\midrule
\textbf{Total questions} & 1,082 & 1,394 \\
\quad Single-page & 494 (45.7\%) & 1,218 (87.4\%) \\
\quad Cross-page & 365 (33.7\%) & 176 (12.6\%) \\
\quad Unanswerable & 223 (20.6\%) & - \\
\midrule
\textbf{Evidence source} & & \\
\quad Pure-text & 305 (35.5\%) & 302 (21.7\%) \\
\quad Layout & 119 (13.9\%) & - \\
\quad Table & 218 (25.4\%) & 573 (41.1\%) \\
\quad Chart & 178 (20.7\%) & 519 (37.2\%) \\
\quad Image & 304 (35.4\%) & - \\
\midrule
\textbf{Avg./Max. question words} & 16.2 / 54 & 35.8 / 108 \\
\textbf{Avg./Max. answer words} & 2.1 / 66 & 23.4 / 174 \\
\bottomrule
\end{tabular}
\end{table}

\subsection{Implementation Details}

\paragraph{Document parsing.} DocFS takes PDF documents as input and first parses them with MinerU \cite{wang2026mineru2, niu2025mineru2, wang2024mineru, he2024opendatalab}, an open-source toolkit for high-quality document content extraction. MinerU produces page-level markdown outputs that preserve text, tables, and images together with their approximate positional information, which serves as the starting point for subsequent preprocessing and file system construction.

\paragraph{Preprocessing.} The raw parsing outputs are further refined in two stages. (1) Section hierarchy reconstruction. We reorganize the flat headings produced by MinerU into a consistent multi-level section tree. This step is carried out with DeepSeek-V4-Flash \cite{deepseekai2026deepseekv4} and DeepSeek-V4.1-Flash, which classify each heading into an appropriate level based on font size, content similarity, and surrounding context. (2) Content correction and image description generation. Parsing artifacts such as malformed table structures and garbled equations are corrected with Qwen-VL-Max \cite{Qwen-VL}, Qwen-VL-Plus \cite{Qwen-VL}, and Qwen3.7-Flash \cite{qwen37}, which are also used to produce concise textual descriptions for images extracted from the document. These descriptions make visual content searchable by text-based retrieval while preserving the original images for fine-grained visual understanding.

\paragraph{Agentic reasoning.} The question-answering agent is built on the DeepAgents \footnote{\url{https://github.com/langchain-ai/deepagents}} framework, a lightweight open-source framework for tool-augmented LLM agents. We retain the official tool set defined by DeepAgents and adapt several tools to the file-system-based setting by modifying or extending their functionality. All backbone models are called through their respective official APIs, and following DocLens we set the temperature to 0.7 for a fair comparison.

\section{Ablation Study Details}
\label{app:ablation}

\textbf{The constructed file system enables DocFS to improve overall performance across different tasks.} Comparing the full DocFS with w/o File System evaluates the overall benefit of the constructed file system over directly using document parsing results. On MMLongBench-Doc, DocFS improves the overall accuracy by 5.0 and 5.3 points and F1 by 5.3 and 4.2 points for Qwen3.5-Plus and DeepSeek-V4.1-Flash, respectively. On FinRAGBench-V, the overall accuracy improves by 3.5 points for Qwen3.5-Plus and 10.4 points for DeepSeek-V4.1-Flash. These results demonstrate that organizing parsed documents into a hierarchical file system, together with structure and image description files, yields consistent overall benefits across different backbone models and datasets. 

\textbf{Reliable preprocessing provides a foundation for subsequent document organization and question answering.} Removing preprocessing which contains section hierarchy reconstruction and equation and table content correction leads to a decrease in the overall accuracy on MMLongBench-Doc by 4.9 points for Qwen3.5-Plus and 11.0 points for DeepSeek-V4.1-Flash, with declines also observed across multiple individual metrics. On FinRAGBench-V, the overall accuracy decreases by 1.7 and 0.7 points for Qwen3.5-Plus and DeepSeek-V4.1-Flash, respectively. These results demonstrate that subsequent file system organization and agent-based question answering still depend on reliable parsing results, and constructing a hierarchical file system without any preprocessing cannot fully compensate for structural and content errors in the initial parsing output. 

\textbf{Structure description files help agents effectively exploit organized structure.} Removing the index and relation files leads to a decrease in the overall accuracy for both backbone models on both datasets. On MMLongBench-Doc, the overall accuracy drops by 10.4 and 5.3 points for DeepSeek-V4.1-Flash and Qwen3.5-Plus, respectively, while on FinRAGBench-V, it decreases by 2.5 points for both models. These results indicate that organizing document contents solely through directory hierarchies and element-type files is insufficient to fully support agents in exploiting the resulting document structure. The index and relation files explicitly provide document navigation and element-level associations, helping agents locate relevant evidence and retrieve related context. Without these structure description files, agents can still access document content, but their ability to explore information through the organized structure is constrained, consequently impairing question-answering performance.

\textbf{Image description files improve the retrieval and utilization of visual evidence.} Removing the textual descriptions of images decreases the overall accuracy on MMLongBench-Doc by 4.7 points for Qwen3.5-Plus and 11.0 points for DeepSeek-V4.1-Flash, and decreases F1 by 5.6 and 9.7 points, respectively. On FinRAGBench-V, the overall accuracy of Qwen3.5-Plus and DeepSeek-V4.1-Flash decreases by 2.5 and 0.4 points, respectively. The degradation is concentrated on questions whose evidence is visual: for Qwen3.5-Plus, the CHA and FIG subsets drop by 11.3 and 7.3 points, and for DeepSeek-V4.1-Flash, the CHA and FIG subsets drop by 15.8 and 10.8 points. On FinRAGBench-V, the CHA and TAB subsets of Qwen3.5-Plus drop by 3.2 and 3.0 points. These findings align with our original motivation for introducing image descriptions: converting key information in visual elements into textual descriptions enables agents to discover potentially relevant images through text-based retrieval and then inspect the original images for fine-grained understanding. The results suggest that image descriptions provide a text-based entry point for retrieving visual elements, helping agents locate and exploit visual evidence and thereby improving overall question-answering performance.


\textbf{Global keyword search is an important access path for locating evidence.} Removing \verb|grep| causes significant drop. On MMLongBench-Doc, the overall accuracy falls by 9.6 and 13.1 points and F1 by 10.7 and 11.9 points for Qwen3.5-Plus and DeepSeek-V4.1-Flash, respectively; on FinRAGBench-V, the overall accuracy falls by 8.3 and 2.5 points. Without global search, the agent has to descend the directory tree level by level and guess which files contain the answer, so evidence scattered across distant sections is easily missed.

\textbf{Directory listing supports structure-aware navigation.} Removing \verb|ls| also lowers the overall accuracy for both backbone models on both benchmarks, by 5.0 and 10.8 points on MMLongBench-Doc and by 1.4 and 0.5 points on FinRAGBench-V for Qwen3.5-Plus and DeepSeek-V4.1-Flash, respectively. Without \verb|ls|, the agent can no longer inspect which sections and files exist at each level of the
hierarchy, and has to reach candidate evidence only through structure description files and keyword search. The decline indicates that browsing the directory structure is a useful complement to content-based search: keyword retrieval locates evidence by matching query terms, while directory listing reveals the available search space itself. Removing either one weakens the agent's ability to exploit the organized structure.

\section{Document Processing Costs}
\label{app:doc_processing_costs}
DocFS builds the file system once per document, covering parsing refinement, organization, and image description generation. Amortized over all questions on a document, this costs 14.5K tokens per question on MMLongBench-Doc and 122.5K on FinRAGBench-V (about 116.2K and 567.3K per document, with 8.0 and 4.6 questions per document on average). For DocLens, the document processing stage (pdf-to-images and parsing) consumes no tokens, so its reported cost, which sums its two reasoning phases, already reflects its full token consumption. Charging DocFS's processing cost fully to each query yields 157.4K/195.4K tokens per question on MMLongBench-Doc (Qwen/DeepSeek) and 307.5K/277.2K on FinRAGBench-V, still below the 217.9K--1,555.9K consumed by DocLens.

\section{Detailed Analysis Figures}
\label{app:detailed_analysis_figures}
This section provides the figures referenced in Section~\ref{sec:detailed_analysis}, including the tool call distribution (Figure~\ref{fig:tool_call_distribution}), the file access distribution (Figure~\ref{fig:read_file_groups}), and the end-to-end reasoning traces (Figure~\ref{fig:case_study}).

\end{document}